\documentclass[twocolumn,aps,prb]{revtex4-1}

\usepackage{amsfonts}
\usepackage{amsmath}
\usepackage{bm}
\usepackage{graphicx}
\usepackage{multirow}

\usepackage{hyperref}

\begin{document}

\title{Supplemental Material}

\author{S. Wimmer, D. K\"odderitzsch, K. Chadova and H. Ebert}

\affiliation{Department Chemie/Phys.\,Chemie, Ludwig-Maximilians-Universit\"at
M\"unchen, Butenandtstrasse 11, 81377 M\"unchen, Germany}

\date{\today}

\maketitle

\subsection*{Symmetry considerations}
\vspace{-0.125cm}
Kleiner \cite{Kle66} introduced a very flexible scheme to 
investigate the symmetry relations for arbitrary response 
functions with the perturbation represented by a vector 
operator. We extended this scheme for the case that the 
response is given by a combination of two vector operators. 
For the spin current density considered here care has to be
 taken furthermore for the fact that one operator is an 
axial and the other one a polar vectorial quantity. With 
this accounted for the structure of the spin conductivity
 as well as spin Nernst conductivity tensors are determined 
applying the restrictions imposed by the cubic point  group. 
Any other lattice symmetry can be treated in the same way. In addition, 
application to magnetically ordered systems runs completely analogously.

\subsection*{ More detailed comparison to previous results}
\begin{figure}[hbt]
  \begin{center}
     \includegraphics[width=\linewidth,clip]{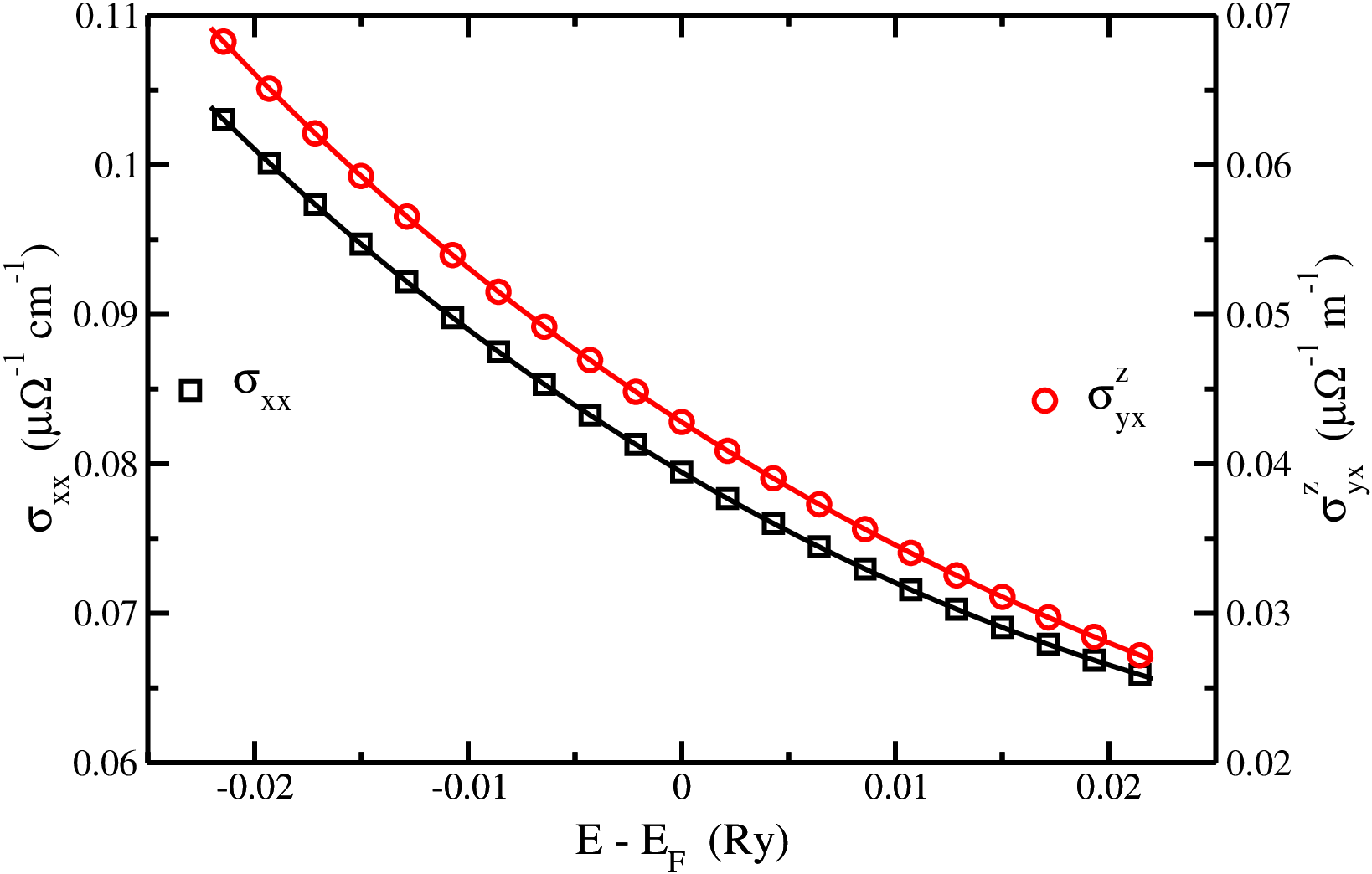}
     \includegraphics[width=\linewidth,clip]{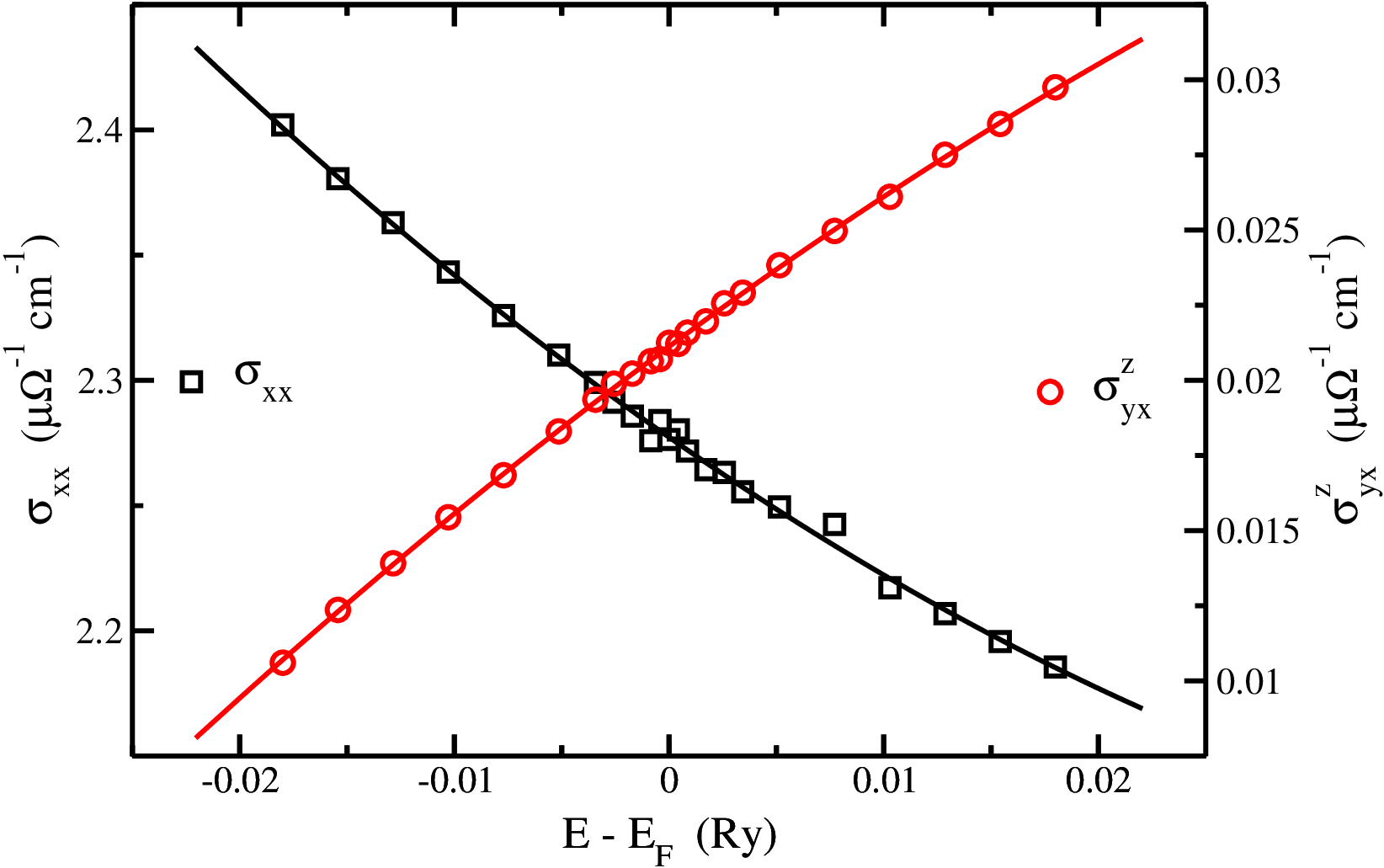}
     \includegraphics[width=\linewidth,clip]{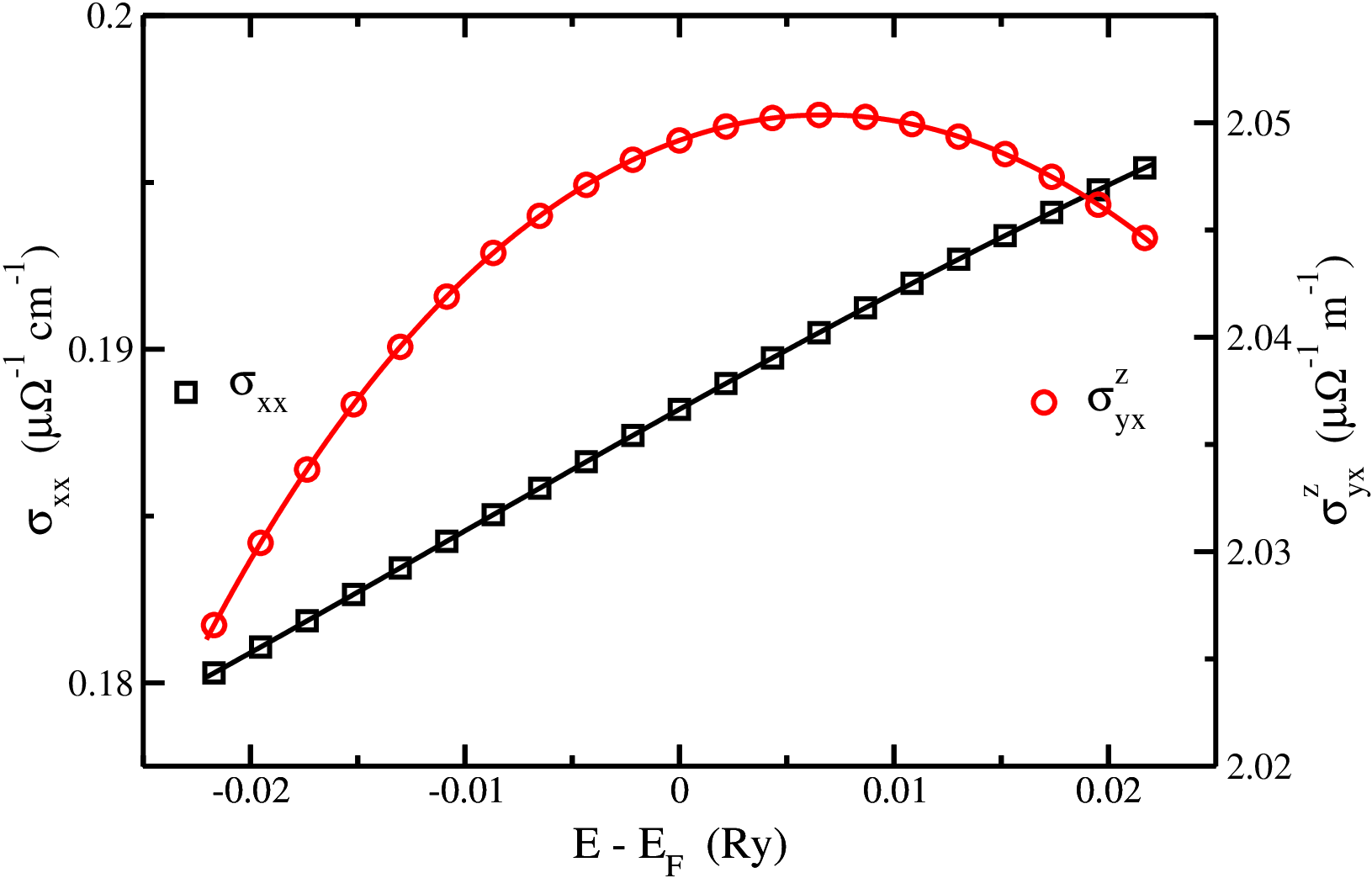}
    \caption{\label{FIG:sofT_VC} (Color online) Energy dependence of the
longitudinal charge and the transverse spin Hall conductivity, $\sigma_{xx}$ and
$\sigma^z_{yx}$ respectively for (from top to bottom) Cu$_{0.99}$Ti$_{0.01}$,
Cu$_{0.99}$Au$_{0.01}$, and 
Cu$_{0.99}$Bi$_{0.01}$.}
\end{center}
\end{figure}

\vspace{-0.125cm}
In Table~I results for several transport properties are compared to those
obtained by Boltzmann transport theory in the work by Tauber 
{\it et al.} (Ref.~\onlinecite{TGFM12}). The following Fig.~\ref{FIG:SxxofT_VC} shows 
the comparison for the charge  Seebeck coefficient $S_{xx}$ \footnote{Which in 
our case has no contributions from transverse response coefficients, unlike 
Eq.~(8) of Ref.~\onlinecite{TGFM12}, since both $L^{cc}$ and $L^{cq}$ in Eq.~(6) 
are diagonal for nonmagnetic cubic crystals. Note that those artificial 
contributions could be shown to be negligibly small.} in more detail, namely 
its temperature dependence.
%
\begin{figure}[hbt]
  \begin{center}
     \includegraphics[width=\linewidth,clip]{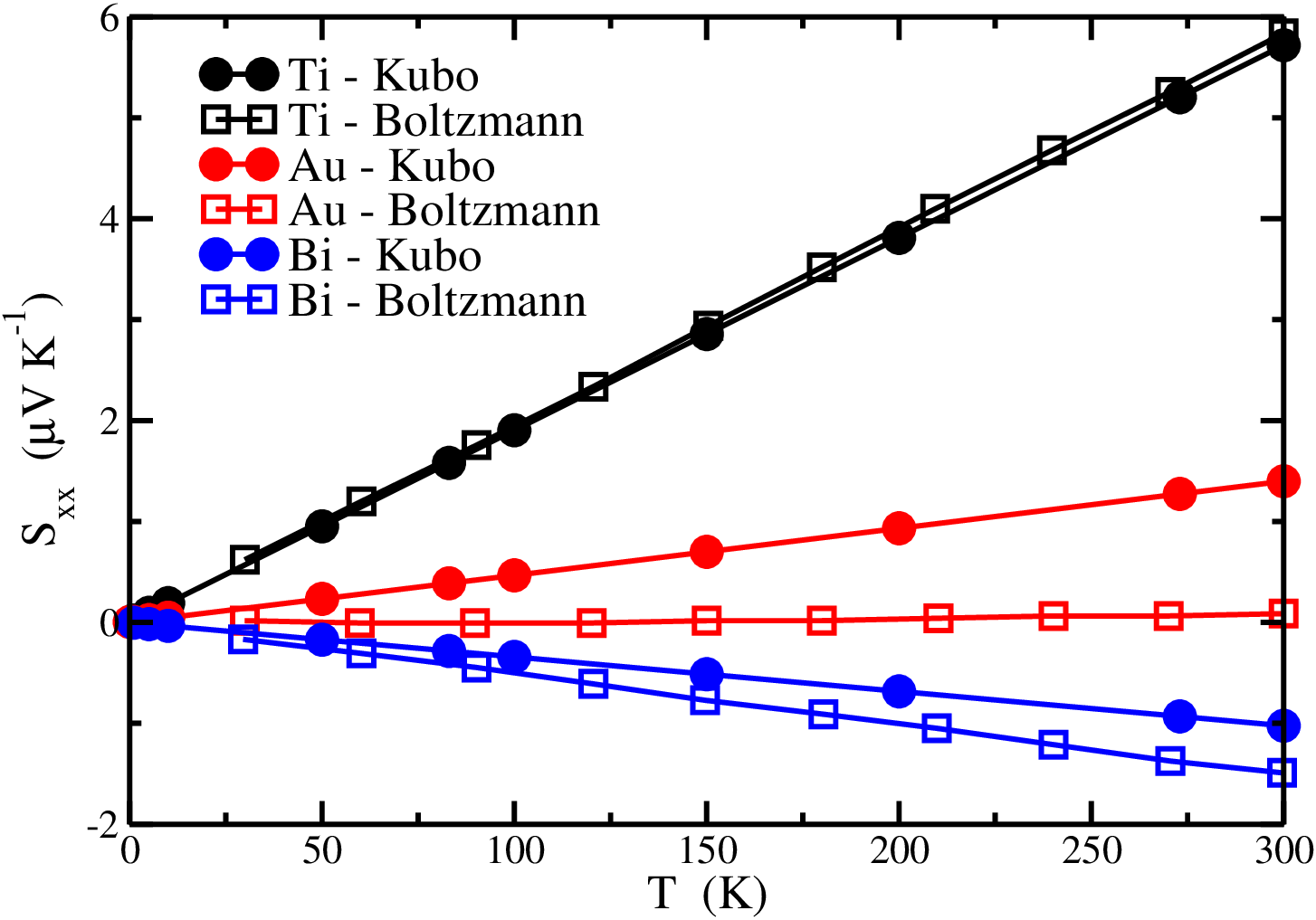}
    \caption{\label{FIG:SxxofT_VC} (Color online) Temperature dependence of the
charge Seebeck
coefficient   $S_{xx}$ in Cu$_{0.99}$Ti$_{0.01}$,   Cu$_{0.99}$Au$_{0.01}$, and 
Cu$_{0.99}$Bi$_{0.01}$
obtained  within Kubo and Boltzmann\cite{TGFM12} transport theory.}
\end{center}
\end{figure}
\vspace{-0.5cm}
%
As a function of temperature all three systems show a linear
increase in magnitude, reflecting the basically linear behavior of the
underlying $\sigma_{xx}(E)$ in the considered energy interval, as can be seen
from Fig.~\ref{FIG:sofT_VC}. The different sign of the Seebeck coefficient in
Cu$_{0.99}$Bi$_{0.01}$ directly traces back to the increase of the longitudinal
conductivity as
a function of energy in the vicinity of the Fermi level, as opposed to
decreasing $\sigma_{xx}(E)$ for Cu$_{0.99}$Ti$_{0.01}$ and
Cu$_{0.99}$Au$_{0.01}$.

%
%

\begin{figure}
  \begin{center}
\includegraphics[width=0.95\linewidth,clip]{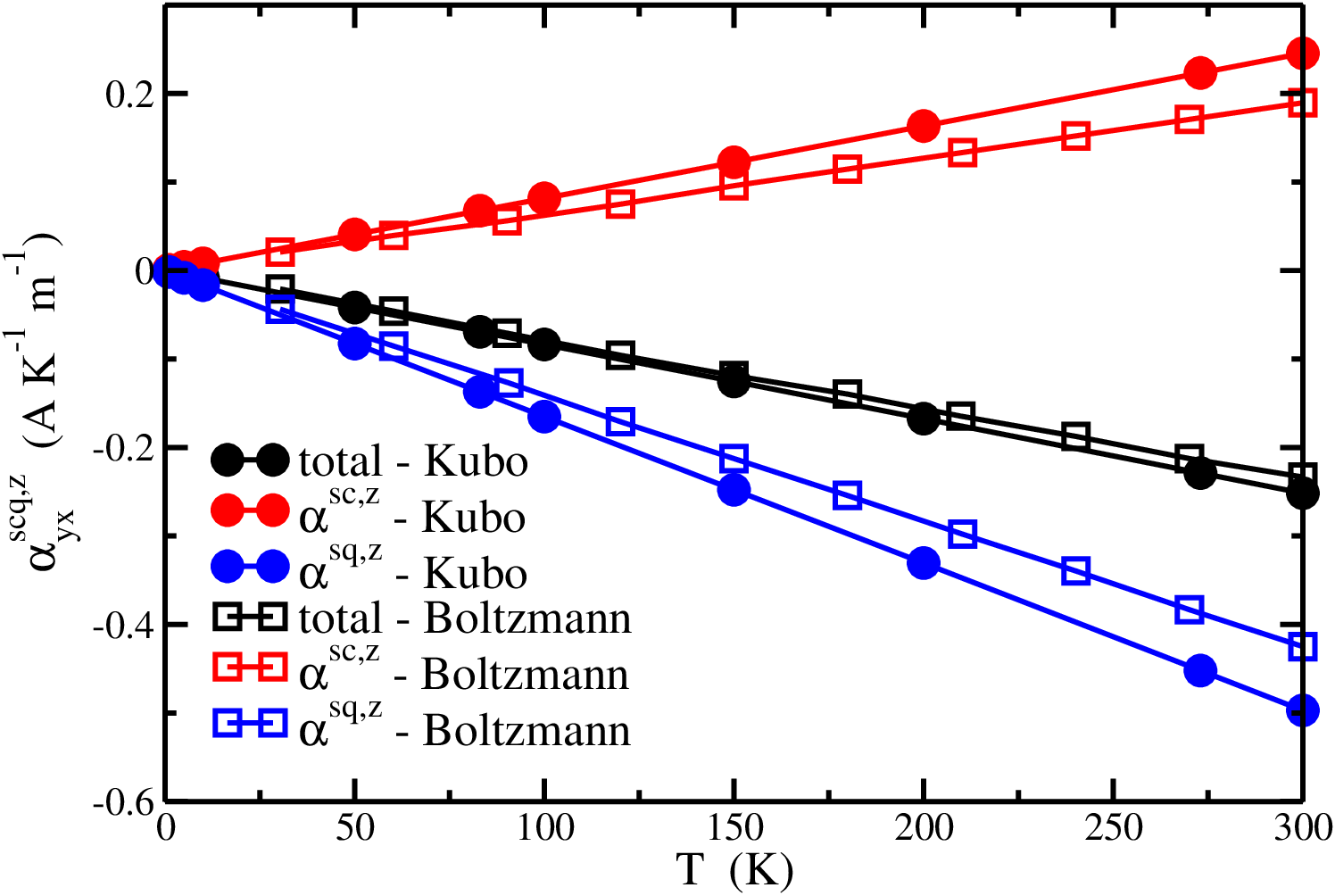}
\includegraphics[width=0.95\linewidth,clip]{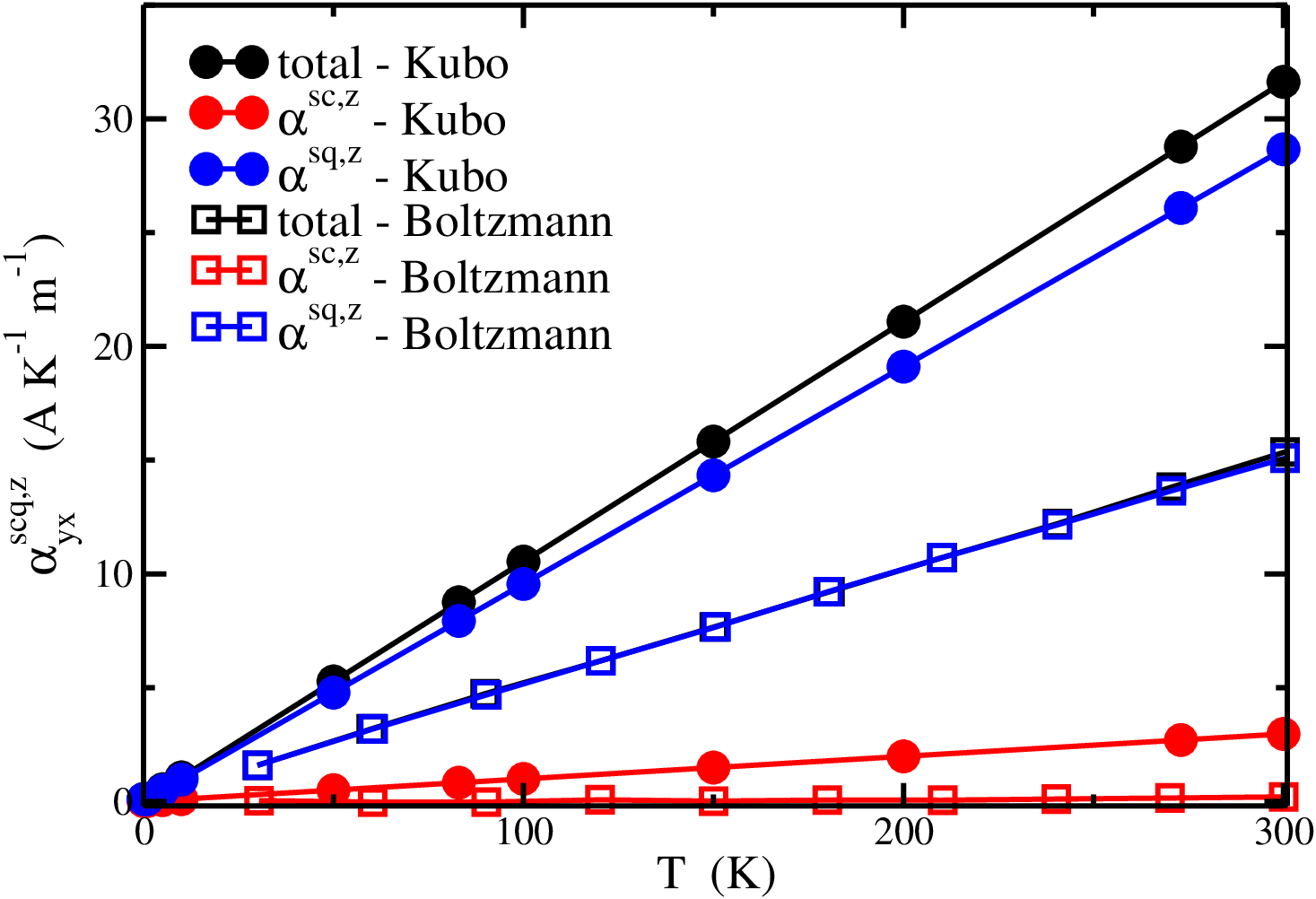}  
\includegraphics[width=0.95\linewidth,clip]{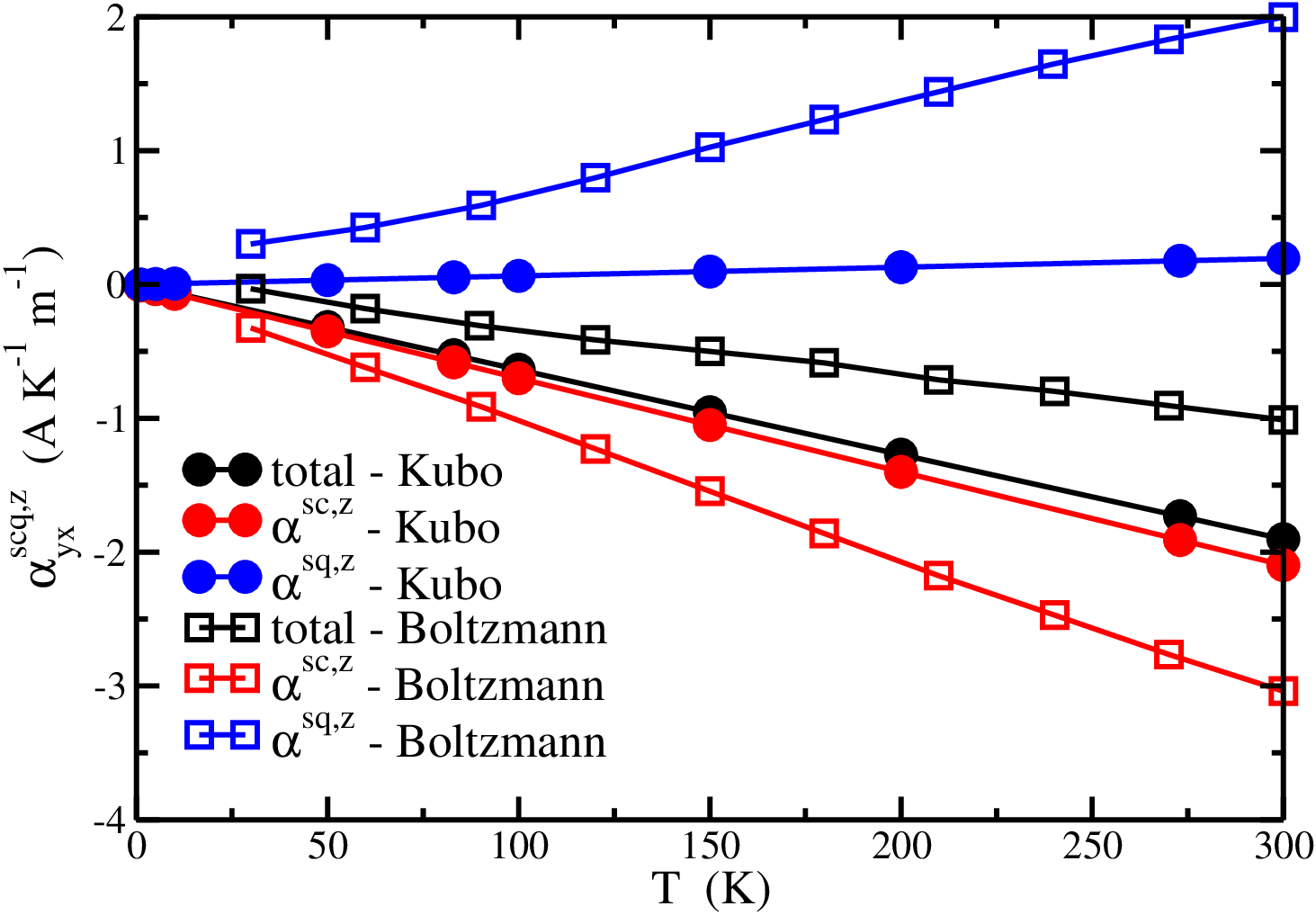}
    \caption{\label{FIG:sigmaSNyx} (Color online) Temperature dependence of the
spin-dependent Nernst conductivity and its
constituents for, from top to bottom,  in Cu$_{0.99}$Ti$_{0.01}$,  
Cu$_{0.99}$Au$_{0.01}$, and 
Cu$_{0.99}$Bi$_{0.01}$   obtained  within Kubo and Boltzmann\cite{TGFM12}
transport theory.}
\end{center}
\end{figure}

A corresponding comparison is shown in Fig.~\ref{FIG:sigmaSNyx} for the total
spin Nernst conductivity 
as well as its individual, electrical and thermal, contributions.
%
%
Again the temperature dependence is approximately linear for all contributions
and all
three systems. The differences in the constitution of the total spin Nernst
conductivity, i.e. the relative magnitudes and signs of the two terms
$\alpha^{sc,z}_{yx}$ and $\alpha^{sq,z}_{yx}$ found in Ref.~\onlinecite{TGFM12}
are
reproduced. Just as for the Seebeck coefficient, magnitude, sign and temperature
dependence of the spin Nernst conductivity (SNC) can be already qualitatively
estimated from the
$\sigma(E)$ curves in Fig.~\ref{FIG:sofT_VC}. Note that for
Cu$_{0.99}$Bi$_{0.01}$ the deviation from linearity is the most prominent, which
results in a moderately non-linear T-dependence of $\alpha^{sq,z}_{yx}$ (nearly
invisible in Fig.~\ref{FIG:sigmaSNyx}, bottom).\\

Results for both quantities show very good agreement for
Cu$_{0.99}$Ti$_{0.01}$
but pronounced deviations for the two other systems containing heavy elements.
For
Cu$_{0.99}$Au$_{0.01}$ this concerns mostly the longitudinal Seebeck coefficient
but also transverse transport properties
while for Cu$_{0.99}$Bi$_{0.01}$ the spin Nernst conductivities,
especially their thermal contributions, deviate. This could possibly be ascribed
to the neglect of spin-flip
 contributions by Tauber {\it et al.},\cite{TGFM12}
 that indeed are expected to increase with the atomic number. But particularly
for longitudinal transport coefficients in Cu$_{0.99}$Au$_{0.01}$ they could be
ruled out to be of significance.\footnote{K. Tauber, D. Fedorov, and I. Mertig 
(private communication).}
 Another possible source for the discrepancies is the the fact that the
Kubo-St\v{r}eda formalism used here gives the full 
conductivities including in particular the intrinsic as 
well as the extrinsic side-jump contributions. These are given
explicitly for Cu(Au) in the dilute limit below and can be shown to be too small
to serve as an explanation. Furthermore, as the two approaches for determining
the electronic structure of the alloy differ insofar as here the CPA is used
whereas in Ref.~\onlinecite{TGFM12} an embedded cluster method has been
employed, differences in the response coefficients are to be expected. In
particular, the energy dependence of the conductivities around the Fermi energy
seems to be
very sensitive. Another possible explanation for the discrepancies is the
description of the spin current density, on the one hand by the use of the
four-component polarization operator\cite{BW48} and on the other hand via the
spin polarization of the Bloch states as outlined in Ref.~\onlinecite{GFZM10}.
Still in all cases the overall agreement concerning
magnitude, sign and temperature dependence is satisfactory.

\subsection*{ Results for diluted Cu-alloy series}
\begin{figure}[hbt]
 \begin{center}
 
\includegraphics[width=\linewidth,clip]{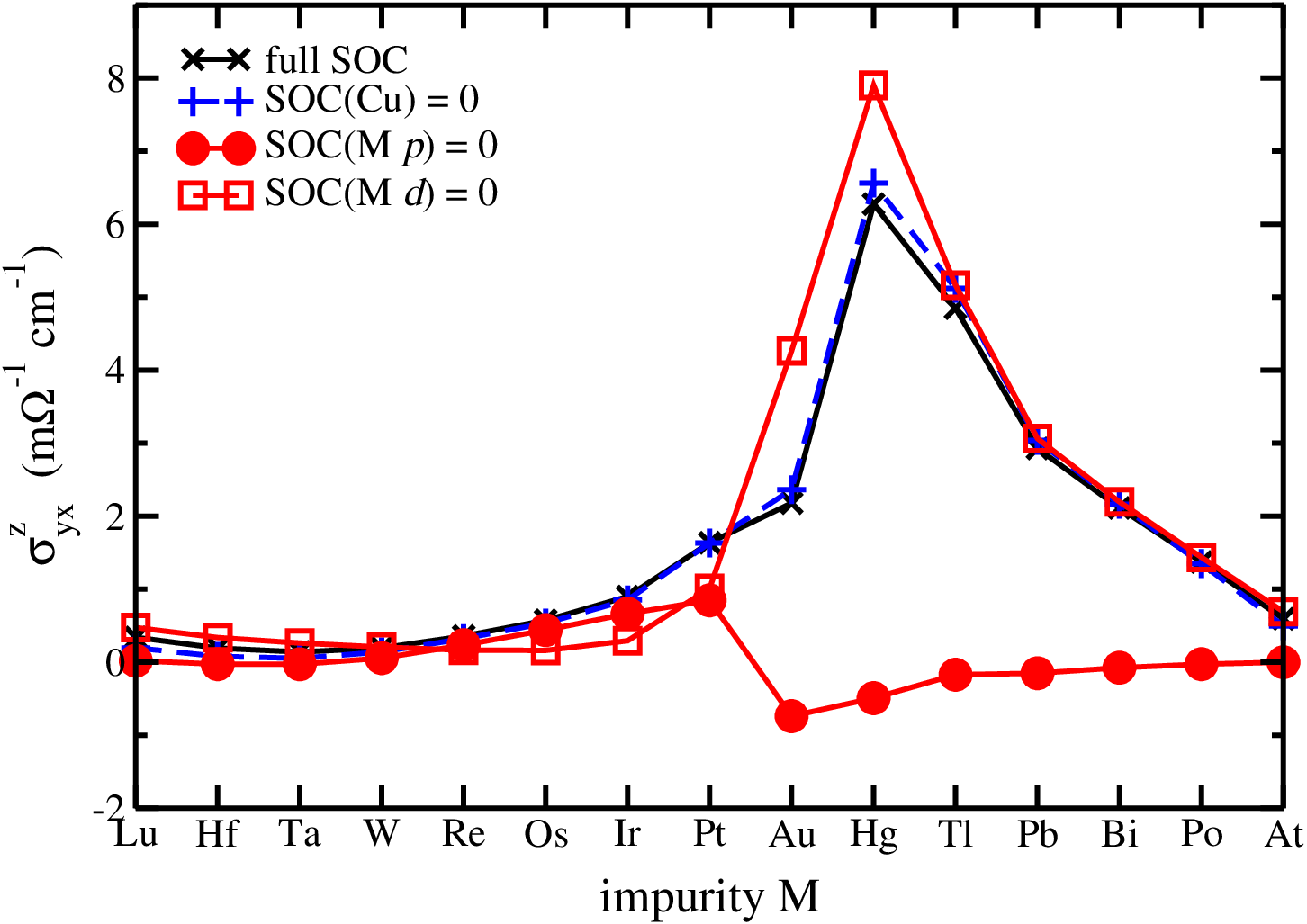}
  \caption{\label{FIG:SOCswitching} (Color online) Spin Hall conductivities
obtained for manipulated spin-orbit coupling (SOC) strength in
Cu$_{0.99}$M$_{0.01}$.}
 \end{center}
\end{figure}

\begin{figure}[hbt]
 \begin{center}
  \includegraphics[width=\linewidth,clip]{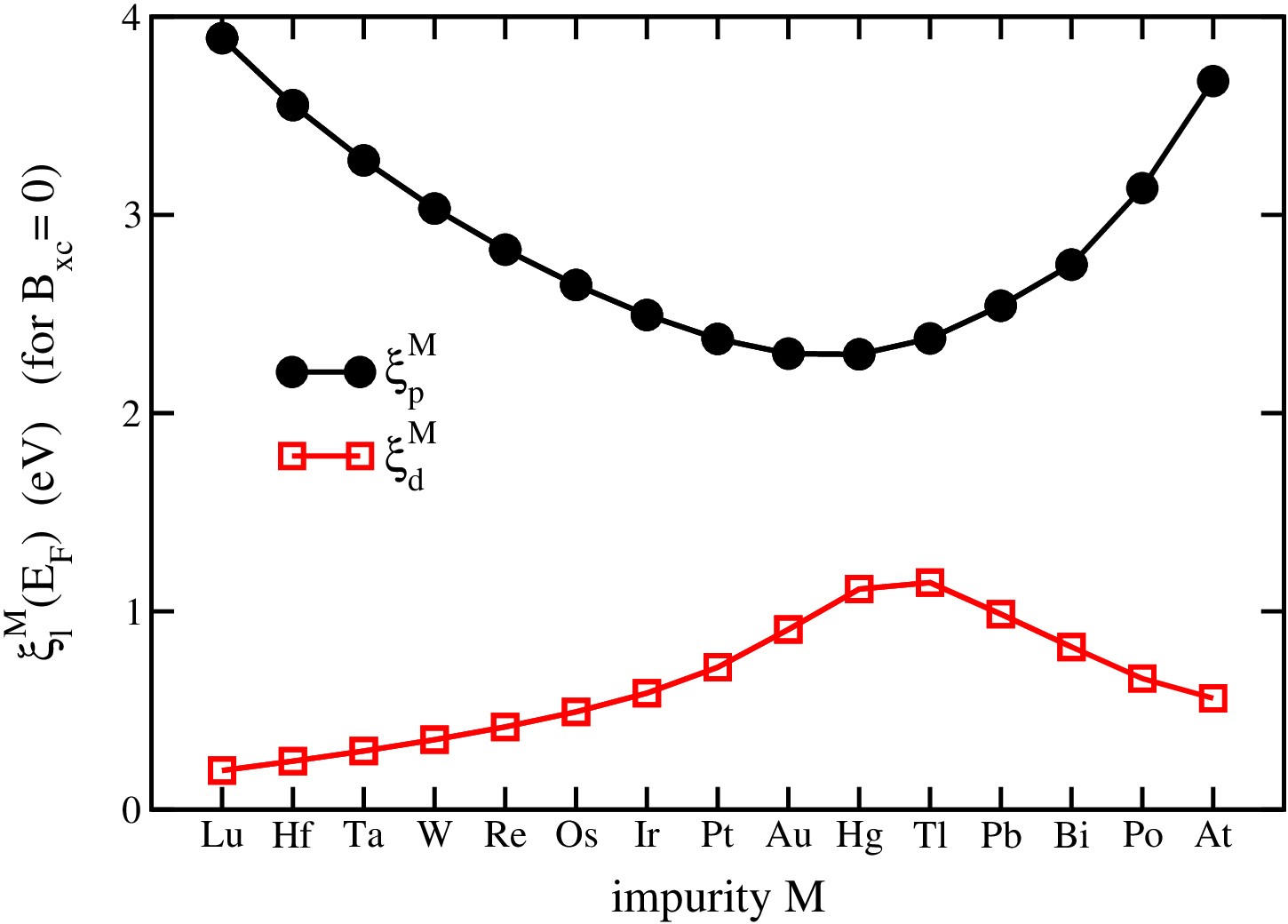}
  \caption{\label{FIG:SOCstrength} (Color online) Spin-orbit coupling strength
of impurity M from Lu to At in
Cu$_{0.99}$M$_{0.01}$}
 \end{center}
\end{figure}

\begin{figure}[hbt]
 \begin{center}
  \includegraphics[width=\linewidth,clip]{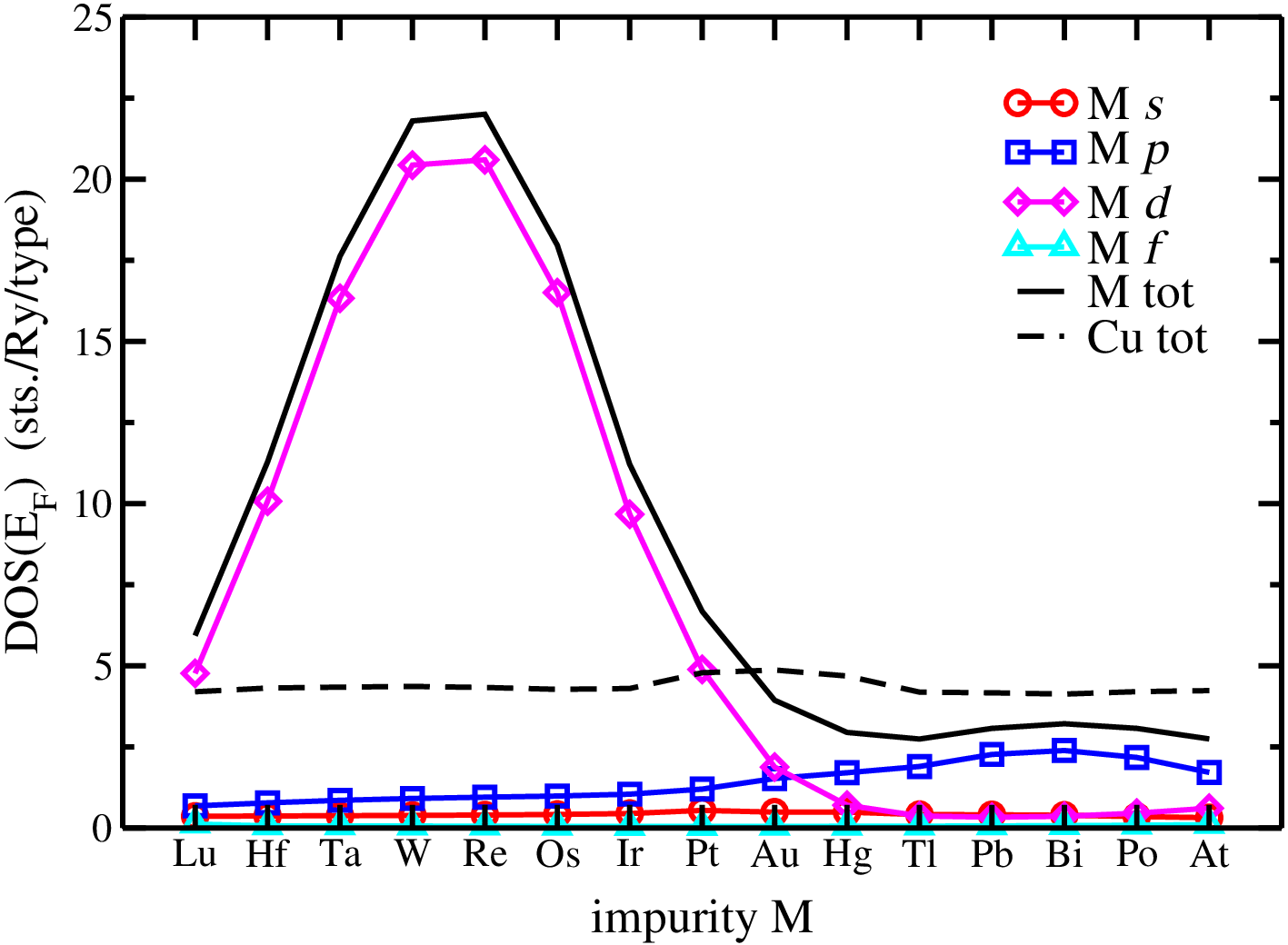}
  \caption{\label{FIG:DOSatEF} (Color online) Density of states (DOS) at the
Fermi energy for impurity M from Lu to At and Cu in
Cu$_{0.99}$M$_{0.01}$. For the former elements the contributions of the
$l$-channels up to $f$ are given in addition.}
 \end{center}
\end{figure}

\vspace{-0.125cm}
The discussion presented on the results of Cu$_{0.99}$Au$_{0.01}$ and 
Cu$_{0.99}$Bi$_{0.01}$ is 
supported by an additional study of the spin Hall conductivity
 for diluted Cu-M alloys. The underlying principles and the used formalism are
outlined in Ref.~\onlinecite{EFVG96}. Suppressing the spin-orbit coupling on the
host element Cu hardly changes 
the spin Hall conductivity as shown in Fig.~\ref{FIG:SOCswitching}.
%
%
 Applying 
the manipulation for the heavy element M on the other hand  
leads to a strong variation of the spin Hall conductivity (SHC),
 in particular for elements around Hg. For this impurity the spin Hall
conductivity with full spin-orbit coupling has the largest value. Performing the
 manipulation individually on the $p$- and $d$-channels shows that the relative
importance of the $p$-channel increases drastically starting from M =
Au.\footnote{Interestingly, also for the elements at the beginning of the
series, M = Lu to W, neglecting $\xi^{M}_p$ drastically reduces the SHC.} Up to
M = Hg also the SOC of the $d$-channel contributes considerably, for Au and Hg
as impurities it even diminishes the SHC. The corresponding spin-orbit coupling
 strength for the elements M (see Fig.~\ref{FIG:SOCstrength}) is found to be
minimal in the $p$-channel at M = Hg, shifted by one to higher atomic numbers,
at Tl, the $d$-channel has its maximum. Furthermore the density of states at the
Fermi energy shows a crossover of the dominance of $d$- to $p$-states between M
= Au and Hg when going from light to heavy elements, as depicted in
Fig.~\ref{FIG:DOSatEF}. All this does not yet provide a full
explanation of the behavior of the SHC as a function of impurity type, it only
hints, by highlighting the necessary ingredients, on the route one has to take
in order to understand the underlying
mechanisms in more detail.
%
%

\subsection*{ Decomposition of spin Hall and spin Nernst conductivity}
\begin{figure}[hbt]
  \begin{center}
   \includegraphics[width=\linewidth,clip]{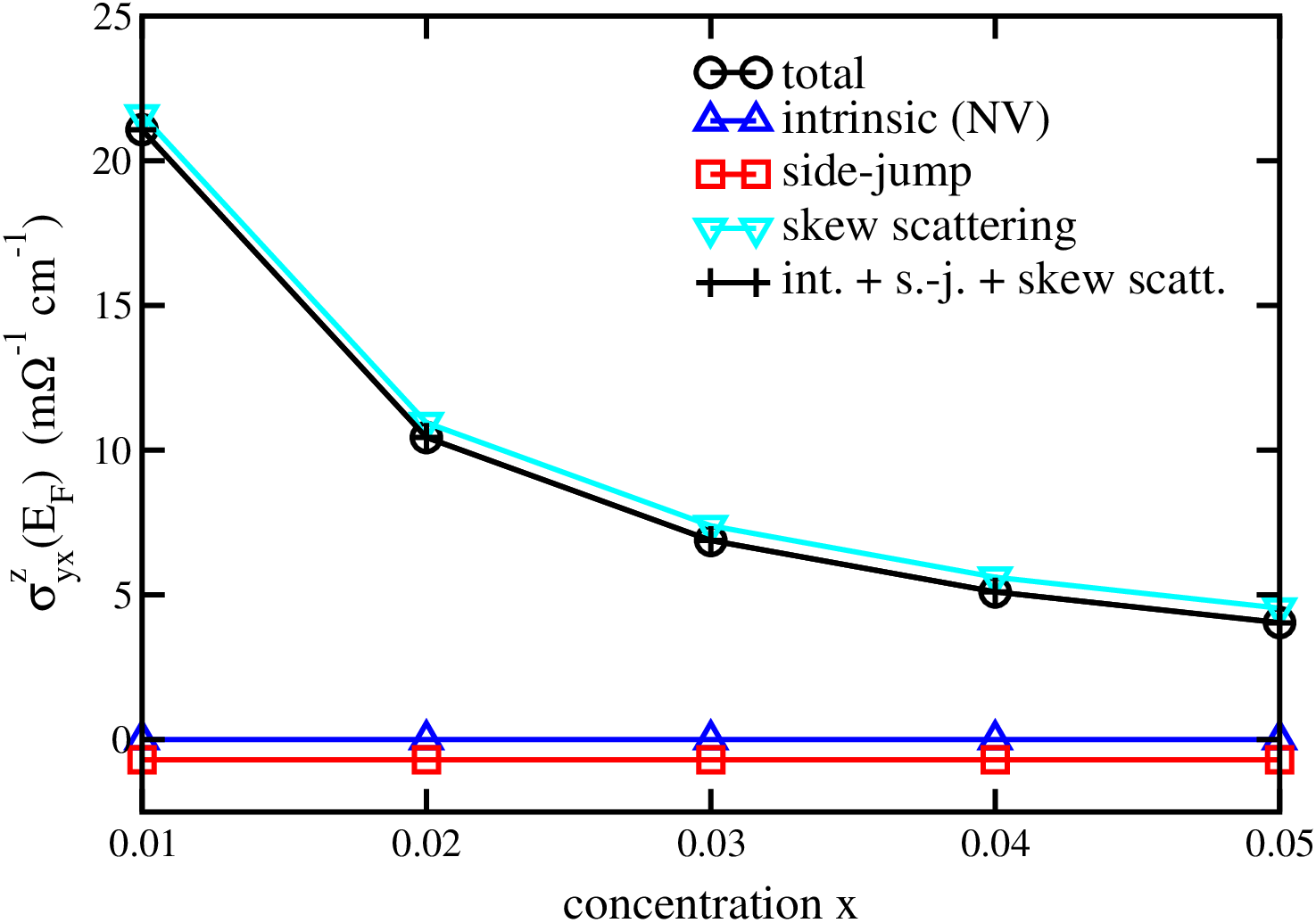}
   \includegraphics[width=\linewidth,clip]{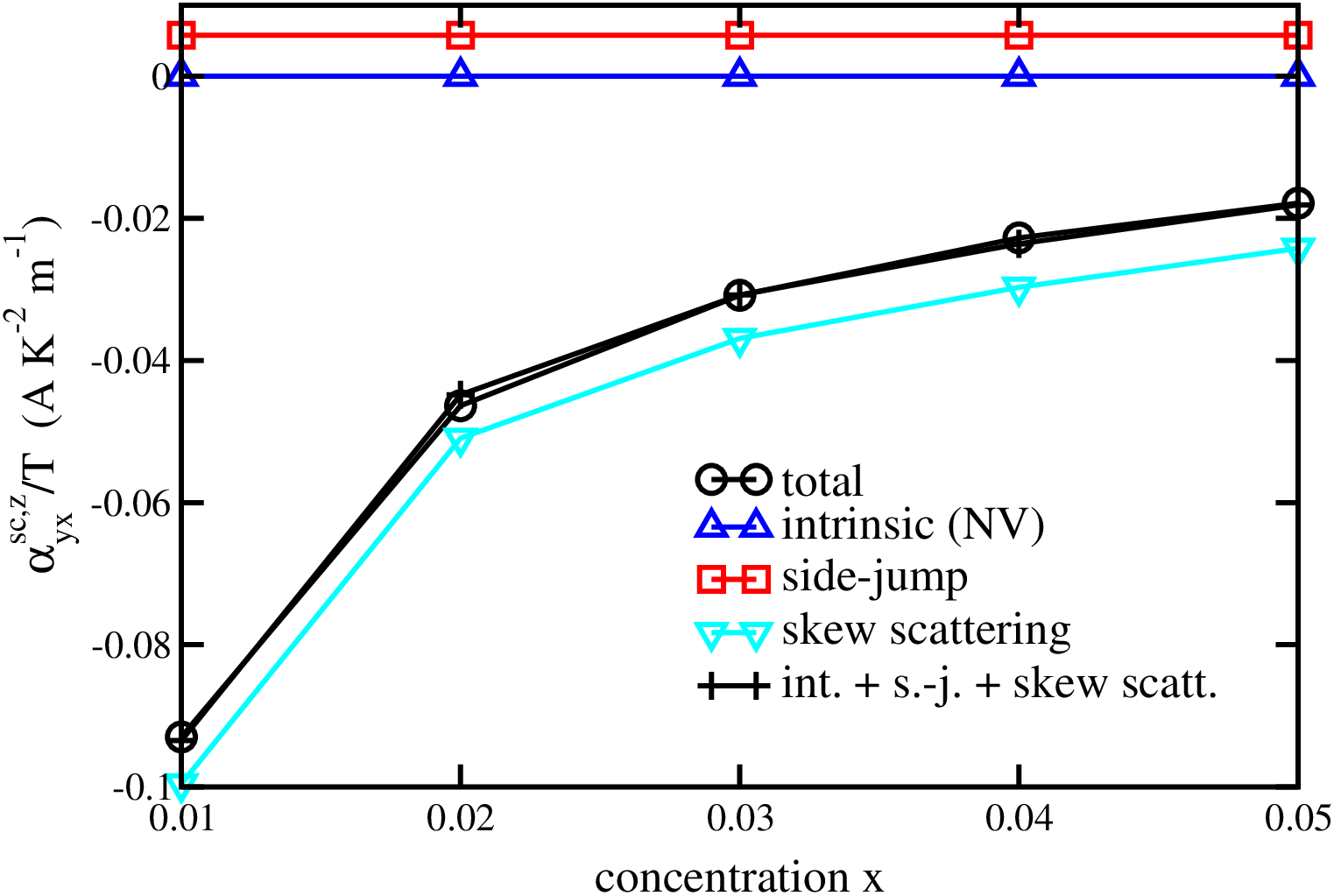}
   \caption{\label{FIG:SHnNCdecomp} (Color online) Decomposition (and
reassembly) of the spin
Hall (top) and spin Nernst conductivity (bottom) on the Cu-rich side of
Au$_x$Cu$_{1-x}$ into intrinsic and
extrinsic (skew scattering and side-jump) contributions.}
  \end{center}
\end{figure}

\vspace{-0.125cm}
Making use of the connection of the vertex corrections
 to the extrinsic contributions to the spin Hall and
 spin Nernst conductivities these have been split
 accordingly into their intrinsic and extrinsic parts. 
For the intrinsic contributions (calculated excluding vertex corrections) in
both cases a linear variation with the concentration is found, as shown in
Fig.~\ref{FIG:SHnNCdecomp}.

Using the scaling behavior\cite{CB01a,OSN08} of the extrinsic contribution 
due to the skew scattering and side-jump mechanisms w.r.t.\
to the longitudinal conductivity a corresponding 
decomposition has been made in addition. Fig.~\ref{FIG:SHnNCdecomp}
shows that the side-jump contributions for both quantities are in the same 
order of magnitude as the intrinsic one and also vary 
only slightly with concentration. The skew scattering 
contribution, on the other hand, gives rise to the diverging 
behavior of both SHC and SNC when approaching the dilute limit.\\

\bibliographystyle{aipnum}

\begin{thebibliography}{10}

\bibitem{Kle66}
W.~H. Kleiner,
\newblock Phys. Rev. {\bf 142}, 318 (1966).

\bibitem{TGFM12}
K.~Tauber, M.~Gradhand, D.~V. Fedorov, and I.~Mertig,
\newblock Phys. Rev. Lett. {\bf 109}, 026601 (2012).

\bibitem{Note1}
Which in our case has no contributions from transverse response coefficients,
  unlike Eq.~(8) of Ref.~\onlinecite{TGFM12}, since both $L^{cc}$ and $L^{cq}$ in
  Eq.~(6) are diagonal for nonmagnetic cubic crystals. Note that those
  artificial contributions could be shown to be negligibly small.

\bibitem{Note2}
K.~Tauber, D.~Fedorov, and I.~Mertig (private communication).

\bibitem{BW48}
V.~Bargmann and E.~P. Wigner,
\newblock Proc. Natl. Acad. Sci. U.S.A. {\bf 34}, 211 (1948).

\bibitem{GFZM10}
M.~Gradhand, D.~V. Fedorov, P.~Zahn, and I.~Mertig,
\newblock Phys. Rev. Lett. {\bf 104}, 186403 (2010).

\bibitem{EFVG96}
H.~Ebert, H.~Freyer, A.~Vernes, and G.-Y. Guo,
\newblock Phys. Rev. B {\bf 53}, 7721 (1996).

\bibitem{Note3}
Interestingly, also for the elements at the beginning of the series, M = Lu to
  W, neglecting $\xi ^{M}_p$ drastically reduces the SHC.

\bibitem{CB01a}
A.~Cr\'epieux and P.~Bruno,
\newblock Phys. Rev. B {\bf 64}, 014416 (2001).

\bibitem{OSN08}
S.~Onoda, N.~Sugimoto, and N.~Nagaosa,
\newblock Phys. Rev. B {\bf 77}, 165103 (2008).

\end{thebibliography}

\end{document}